\documentstyle[12pt]{article}
\begin{document}
\newcommand{\text}{\mbox}

\begin{center}
\smallskip
\-{\LARGE On q-deformed P\"{o}schl-Teller potentials}
\end{center}

\bigskip%

\begin{center}
{\bf Abilio De Freitas}$\footnote{%
Present address: UCLA,\ Physics Deparment, Box 951361. Los Angeles,
CA-90095-1361.}${\bf \ and Sebasti\'{a}n Salam\'{o}}$\footnote{%
E-mail adress: ssalamo@fis.usb.ve}$

Universidad Sim\'{o}n Bol\'{\i}var, Departamento de F\'{\i}sica, Apartado Postal 89000

Caracas,Venezuela

\medskip
\end{center}

\bigskip

\begin{center}
{\bf Abstract}
\end{center}

A simple algebraic technique is developed to obtain deformed energy spectra
for the P\"{o}schl-Teller potentials.

\medskip


\bigskip%

\section{Introduction}

\smallskip

There has been an increasing interest in quantum deformed systems during the
past decade [1]. The use of q-deformed algebras has been seen as a possible
generalization of the usual algebrization of the Schr\"{o}dinger equation
with the use of Lie algebras. There are several approaches to this problem.
Bonatsos et al. [2] have studied deformed harmonic and anharmonic
oscillators as a possible description of vibrational spectra of diatomic
molecules. More recently, Cooper et al [3] have studied the deformation of
the Morse potential using supersymmetric quantum mechanics (SSQM) and a
group approach. It is interesting to notice that almost all physical
problems for which a deformation has been carried out belong to the class of
potentials related to confluent hypergeometric functions, i.e., Coulomb,
harmonic oscillator and Morse potentials, the only exception seems the case
treated in [4].

In this paper, following the ideas developed in [3], we deal with two
potentials whose solutions are hypergeometric functions, the
P\"{o}schl-Teller I and II, using only the spectrum generating algebra
associated with these potentials. No reference to SSQM is used.

Recently in [5] it was shown that the hypergeometric Natanzon potentials
[6],\ $V_{N\text{ }}$ - those for which the Schr\"{o}dinger equation can be
transformed to an hypergeometric function- can be solved algebraically by
means of the SO(2,1) algebra. This is an essential point for what is going
to be done later, so it is convenient to make a brief review of the subject.
The basic assumptions of this approach are: a) a two variable realization of
SO(2,1). b) The Schr\"{o}dinger equation can be written in terms of the
Casimir operator of the algebra\ $\ C$,\ as $\ \left[ H-E\right] $ $\Psi
(r,y)=G(r)[C-c]\Psi (r,y)\ $, where\ \ $c\ $ is the eigenvalue of\ $C$ , $H\
$ the Hamiltonian and\ $E$ \ the corresponding eigenvalue. $G(r)$ is a
function fixed by consistency, and\ c) The eigenfunctions of the Hamiltonian
have the form $\Psi (r,y)=\exp (imy)\Phi (r)$.

The hypergeometric Natanzon potentials are given by:

\begin{eqnarray}
V_{N} &=&\frac{1}{R}(\ f\ z(r)^{2}-(\ h_{0}-h_{1}+f\ )\ z(r)+h_{0}+1) \\
&&+\frac{z(r)^{2}(1-z(r))^{2}}{R^{2}}\left[ a+\frac{a+(c_{1}-c_{0})(2\
z(r)-1)}{z(r)\ (z(r)-1)}-\frac{5\Delta }{4R}\right]  \nonumber
\end{eqnarray}
where

\begin{equation}
\Delta =\tau ^{2}-4ac_{0}\;,\;\tau =c_{1}-c_{0}-a\;,\;R=a\ z(r)^{2}+\tau \
z(r)+c_{0}
\end{equation}
\ The constants $\ a,\ c_{0},\ c_{1},\ h_{0},\ h_{1}\ $and $\ f$ \ are
called Natanzon parameters. The function\ $\ z(r)\ $ must satisfy

\begin{equation}
\frac{dz(r)}{dr}=\frac{2z(r)(1-z(r))}{\sqrt{R}}
\end{equation}
We follow the notation of [7].

The generators of the SO(2,1) algebra:\ $J_{1}$, $J_{2}$ and $\ J_{0}\ $
satisfy the usual commutation relations: $\ [J_{0},J_{1}]=iJ_{2\ }$,$\
[J_{2},J_{0}]=iJ_{1\ }$,$\ [J_{1},J_{2}]=-iJ_{0\ }$ , as usual we define $\
J_{\pm }=J_{1}\pm i\,J_{2}$. The Casimir operator $C$ is given by $\
C=J_{0}(J_{0}\pm 1)-J_{\mp }J_{\pm \ }$.The generators are then given by

\begin{equation}
\exp (_{\mp }iy)\,J_{\pm }=\pm \left( \frac{\,z(r)^{1/2}(z(r)-1)}{%
z(r)^{^{\prime }}}\right) \frac{\partial }{\partial r}-\left( \frac{i}{2}%
\frac{(z(r)+1)}{\sqrt{z(r)}}\right) \frac{\partial }{\partial y}
\end{equation}

\[
+\frac{(z(r)-1)}{2}\left[ \frac{(p\mp 1)}{\sqrt{z(r)}}-\frac{\sqrt{z(r)\ }%
z(r)^{^{\prime }\,^{\prime }}}{z(r)^{^{\prime }\,2}}\right]
\]

\begin{equation}
J_{0}=-i\frac{\partial }{\partial y}
\end{equation}
where $\ z(r)^{^{\prime }}=\frac{dz(r)}{dr}$ \ and\ $\ p$ is a function of
the Natanzon parameters and they generally depend on the energy of the
system. The Casimir operator turns out to be

\begin{equation}
C=(z(r)-1)^{2}\left[ \frac{z(r)}{z(r)^{^{\prime }\,2}}\frac{\partial ^{2}}{%
\partial r^{2}}+\frac{i}{4\sqrt{z(r)}}\frac{\partial ^{2}}{\partial y^{2}}+%
\frac{i\ p\ (z(r)+1)}{2\ (z(r)-1)\ z(r)}\frac{\partial ^{{}}}{\partial y}%
\right]
\end{equation}
\newline
\newline

\[
+(z(r)-1)^{2}\left[ \frac{z(r)\ z(r)^{^{\prime \prime \prime }}}{2\
z(r)^{^{\prime }\ 2}}-\frac{3\ z(r)\ z(r)^{^{\prime \prime \ }2}}{4\
z(r)^{^{\prime }\ 4}}-\frac{(p^{2}-1)}{4\ z(r)}\right]
\]

The eigenvalues of the compact generator, (5) are known to be

\begin{equation}
m=\nu +\frac{1}{2}+\sqrt{\ c+\frac{1}{4}\ \ },\qquad \nu =0,1,...
\end{equation}
and the energy spectra is given by

\begin{equation}
2\nu +1=\alpha (\nu )-\beta (\nu )-\delta (\nu )
\end{equation}
where

\begin{equation}
\alpha (\nu )=\sqrt{-aE(\nu )+f+1},\ \beta (\nu )=\sqrt{-c_{0}E(\nu )+h_{0}+1%
},\ \delta (\nu )=\sqrt{-c_{1}E(\nu )+h_{1}+1}
\end{equation}

The last relevant relation for this work is the connection between the
eigenvalues of the Casimir operator, $c$ , with $E(\nu )$ \ and the Natanzon
parameters:

\begin{equation}
\sqrt{-c_{1}E(\nu )+h_{1}+1}=\sqrt{4\ c+1}
\end{equation}

\bigskip%

\section{Deformed potentials}

\smallskip

We proceed in a similar way as in reference [3] to obtain the deformed
energy spectra of the P\"{o}schl - Teller potentials. For this purpose we
consider the following bosonic realization of SO(2,1) [8]

\begin{equation}
J_{+}=a^{\dagger }b^{\dagger }\quad ,\ J_{-}=ab\quad ,\ J_{0}=\frac{1}{2}\
\left( a^{\dagger }a+b^{\dagger }b+1\right)
\end{equation}
the operators $\ a$, $a^{\dagger }$, $b$ and $b^{\dagger }$ have the usual
commutation relations : $\left[ a,a^{\dagger }\right] =\left[ b,b^{\dagger
}\right] =1$, while the other commutators are zero. The basis for this
realization is given by

\begin{equation}
|j\ m\rangle =\frac{a^{\dagger m-j}b^{\dagger m+j-1}}{\sqrt{(m-j)!\ (m+j-1)!}%
}\ |0\rangle
\end{equation}
in this basis, the operators defined in (11) satisfy the following relations

\begin{equation}
J_{\pm }\ |j\ m\rangle =\sqrt{(m\mp j\pm 1)\ (m\pm j)}\ |j\ m\pm 1\rangle
\end{equation}

\begin{equation}
J_{0}\ |j\ m\rangle =m\ |j\ m\rangle
\end{equation}

The Casimir operator has eigenvalues: $j\ (j-1)$ and $\ j\ \ $is assumed to
be positive, $m$\ takes the values: $m=j+\nu $, $\nu =0,1..$.

The deformed states are defined as [9]

\begin{equation}
|j\ m\rangle _{q}=\frac{a^{\dagger m-j}b^{\dagger m+j-1}}{\sqrt{[m-j]!\
[m+j-1]!}}\ |0\rangle _{q}
\end{equation}
where\ $\ [n]!=[1][2]...,[n]$ , with

\begin{equation}
\lbrack n]=\frac{q^{n}-q^{-n}}{q-q^{-1}}
\end{equation}
and the $q$-boson operators satisfy the following commutations relations

\begin{equation}
aa^{\dagger }-qa^{\dagger }a=q^{-N_{a}}\ ,\ [N_{a}\ ,a^{\dagger
}]=a^{\dagger }\ ,\ [N_{a}\ ,a]=-a
\end{equation}
with similar relations for the $b$ operators. The relations given in (13)
are transformed into the following ones:

\begin{equation}
J_{\pm }\ |j\ m\rangle _{q}=\sqrt{[m\pm j]\ [m\mp j\pm 1]}\ \ |j\ m\pm
1\rangle _{q}
\end{equation}

\begin{equation}
J_{0}\ |j\ m\rangle _{q}=m\ |j\ m\rangle _{q}
\end{equation}
Then the ladder operators defined in (13) satisfy the SO(2,1)$_{q}$ algebra

\begin{equation}
\lbrack J_{0},\ J_{\pm }]=\pm \ J_{0}\ ,\quad [J_{+},\ J_{-}]=-[2\
J_{0}]\quad
\end{equation}

We can now find the deformed energy spectrum in the following way: First we
build up a bosonic representation of the Hamiltonian for the system in
consideration- $H_{b}\ $-in such a way that its spectrum coincides with the
one obtained from (8). The next step is a deformation of $H_{b}$ and the
calculation of its spectrum. Let us consider a few examples, we follow the
notation of [10].

\smallskip

\noindent1) {\bf P\"{o}schl-Teller I potential}

This potential is defined by

\begin{equation}
V=-(A+B)^{2}+A(A-\alpha )\sec ^{2}(\alpha x)+B(B-\alpha )\csc ^{2}(\alpha x)
\end{equation}
it is easily seen that the Natanzon parameters for this potential are given
by

\begin{eqnarray}
a &=&0,\ c_{0}=0,\ c_{1}=-\frac{1}{\alpha ^{2}},\ f=\frac{(2A+\alpha )\
(2A-3\alpha )}{4\alpha ^{2}} \\
\ h_{0} &=&\frac{(2B+\alpha )\ (2B-3\alpha )}{4\alpha ^{2}},\ h_{1}=\frac{%
(A+B+\alpha )\ (A+B-\alpha )}{\alpha ^{2}}  \nonumber
\end{eqnarray}
the function\ $z(x)$ is obtained from (3) and it becomes\ $z(x)=-\tan
^{2}(\alpha x)$. After a careful study of the signs of the square roots
occurring in (8) one finds that the energy spectrum of this system is
\begin{equation}
E(\nu )=4\upsilon \alpha (\upsilon \alpha +A+B)\ \ ,\quad \nu =0,1,...
\end{equation}
>From (10) we obtain for $c$, the eigenvalue of the Casimir operator,

\begin{equation}
c=\frac{1}{4\alpha ^{2}}\left( (A+B+2\nu \alpha )^{2}-\alpha ^{2}\right)
\end{equation}
from (7) $m$ is found to be

\begin{equation}
m=2\nu +\frac{1}{2\alpha }(A+B+\alpha )=2\nu +\lambda
\end{equation}
and from (24)

\begin{equation}
j=\nu +\frac{1}{2\alpha }(A+B+\alpha )=\nu +\lambda
\end{equation}
where the parameter $\lambda $ is defined as : $\lambda =\frac{1}{2\alpha }%
(A+B+\alpha )$. Denoting the adimensional energy of the system by $e(\nu )$,
we have from (23)

\begin{equation}
e(\nu ,\lambda )=4\nu (\nu +2\lambda -1)
\end{equation}
For the values of\ \ $j$ \ and\ $\ m$ given in (25-26), the corresponding
state will be written as $|\nu \ \lambda \rangle $, then it is easily seen
that: $\ a^{\dagger }a\ |\nu \ \lambda \rangle =\nu \ |\nu \ \lambda \rangle
$ \ and $\ b^{\dagger }b\ |\nu \ \lambda \rangle =(2\lambda +\nu -1)\ |\nu \
\lambda \rangle $. With these results we can build a bosonic representation
of the Hamiltonian operator $H_{pt1}$ for this system as follows

\begin{equation}
H_{pt1}=a^{\dagger }a\ (b^{\dagger }b-2a^{\dagger }a)
\end{equation}
then we have the expected result

\begin{equation}
H_{pt1}\ |\nu \ \lambda \rangle =e(\nu ,\lambda )\ |\nu \ \lambda \rangle
\end{equation}
Keeping the same form for the deformed Hamiltonian as the one given in (28),
we obtain for the deformed version of (29)

\begin{equation}
H_{dpt1}\ |\nu \ \lambda \rangle _{q}=e(\nu ,\lambda )_{q}\ |\nu \ \lambda
\rangle _{q}
\end{equation}
with

\begin{equation}
e(\nu ,\lambda )_{q}=4\ [\nu ]\ (\ [2\lambda +3\nu -1]-2[\nu ]\ )
\end{equation}
and $H_{dpt1}\equiv a_{q}^{\dagger }a_{q}(b_{q}^{\dagger
}b_{q}-2a_{q}^{\dagger }a_{q})$. The expression (31) can be considered as
the deformed energy spectrum of the system, its limit as $q\rightarrow 1$
agrees with the undeformed energy given in (23).

\smallskip

\smallskip

\noindent2) {\bf P\"{o}schl-Teller II potential}

It is defined by

\begin{equation}
V=(A-B)^{2}-A(A+\alpha )\text{sech}^{2}(\alpha r)^{{}}+B(B-\alpha )\text{csch%
}^{2}(\alpha r)
\end{equation}
The Natanzon parameters are

\begin{eqnarray}
a &=&0,\ c_{0}=0,\ c_{1}=\frac{1}{\alpha ^{2}},\ f=\frac{(2A-\alpha
)(2A+3\alpha )}{4\alpha ^{2}} \\
\ h_{0} &=&\frac{(2B+\alpha )(2B-3\alpha )}{4\alpha ^{2}},\ h_{1}=\frac{%
(A-B+\alpha )(A-B+\alpha )}{\alpha ^{2}}  \nonumber
\end{eqnarray}
With this set of parameters one obtains from (3,8,9) ,\ $z(r)=\tanh
^{2}(\alpha r){}\ $and for the energy spectra

\begin{equation}
E(\nu )=-4\nu \alpha \ (\nu \alpha -A+B)\ \ ,\quad \nu =0,1,...
\end{equation}
In this case $c$, $m$ and\ $\ j$ are given by

\begin{eqnarray}
c &=&\frac{1}{4\alpha ^{2}}\ \left( (A-B-2\nu \alpha )^{2}-\alpha ^{2}\right)
\nonumber \\
m &=&\frac{1}{2\alpha }\ (A-B+\alpha )\equiv \lambda \\
j &=&\frac{1}{2\alpha }\ (A-B+\alpha )-\nu =\lambda -\nu  \nonumber
\end{eqnarray}
while $e(\nu ,\lambda )$ is now

\begin{equation}
e(\nu ,\lambda )=4\nu \ (2\lambda -\nu -1)
\end{equation}
The corresponding Hamiltonian operator is a simple one

\begin{equation}
H_{pt2}=4\ J_{+}J_{-}
\end{equation}
For the deformed energy spectrum we find

\begin{equation}
e(\nu ,\lambda )_{q}=4\ [\nu ]\ [2\lambda -\nu -1]
\end{equation}
One easily see that the limiting case of (38) agrees with the result given
in (36).

At this point we would like to make a few remarks. One may ask why we are
considering these two potentials and not analyzing the rest of the shape
invariant hypergeometric potentials listed for example in [10]. There are
several reasons. We first notice that the energy spectra of these class of
potentials have two main features, one is that there is a sub - class of
potentials whose their energy spectra is a quadratic expression involving $%
\nu $ while the spectrum of the other ones is a ratio of quadratic
polynomials in $\nu $. In the first case, there are only two cases where two
Natanzon parameters occurs in the energy spectra, these are the cases
treated here, in the remaining two cases only one parameter appears. For the
latter cases one can assume that one parameter is proportional to the other
and the results are not so simple as the cases treated in this paper. For
the remaining set of potentials, the bosonization is not so clear since one
has to get ratios of quadratics polynomials in $\nu $ .

\smallskip

Finally we would like to mention that a polynomial deformation, in the sense
of Delbecq et al [11], can be done in an simple way for the
P\"{o}schl-Teller II potential. This is so because of the simple structure
of $\ H_{pt2}$ \ given in (37). The algebra that we use is the ${\cal A}%
_{q}^{-}$ $(2,1)$ , and we use the $D_{q}^{(+)}$ representation which is
bounded below. These algebra satisfy the following commutation relations

\[
\lbrack J_{0\ },\ J_{+}]=(1+(1-q)\ J_{0})\ J_{+}
\]

\begin{equation}
\lbrack J_{0\ },\ J_{+}]=-J_{-}\ (1+(1-q)\ J_{0})\
\end{equation}

\[
\lbrack J_{+\ },\ J_{-}]=-2\ J_{0}\ (1+(1-q)\ J_{0})
\]
For this algebra the Casimir operator is given by

\begin{equation}
C_{q}=J_{+}\ J_{-}-\frac{2}{1+q}\ (q\ J_{0}-1)\ J_{0}
\end{equation}
The eigenvalues of the Casimir operator are written as follows

\begin{equation}
c_{q}=-\frac{2}{(1+q)}\phi \ (\phi +1)
\end{equation}
with $\ \phi =q\ j<0$. The eigenvalues of $\ J_{0\ \ }$are

\begin{equation}
m_{q}=-j\ q^{-\upsilon }-\frac{1-q^{-\upsilon }}{1-q},\quad \nu =0,1,...
\end{equation}

Notice that$\ m_{q}\rightarrow m$ in the limit $q\rightarrow 1$, in other
words $D_{q}^{(+)}\rightarrow D^{(+)}\ $ which is the representation that is
used in the standard algebraic description of the Natanzon potentials.

As before, the deformed Hamiltonian is assumed to have the same form given
in (37), then from (40) we can write this deformation as
\begin{equation}
H_{pt2}^{^{\prime }}=4\ (C_{q}+\frac{2}{1+q}\ (q\ J_{0}-1)\ J_{0})
\end{equation}
The eigenvalues of $\ H_{pt2}^{^{\prime }}$ are thus given by

\begin{equation}
e^{^{\prime }}(\nu ,\lambda )_{q}=\frac{8}{1+q}\ (\ m_{q}\ (qm_{q}-1)-qj\
(qj+1)\ )
\end{equation}
where in the expression for\ $m_{q}$ we must use the results given in (35),
namely\ $\ m=\lambda $ \ and\ \ $j=\lambda -\nu $ .This is the deformed
energy spectra for deformation in consideration. It is easy to prove that in
the $\ q\rightarrow 1\ $ limit one recovers the result given in (36).

\bigskip%

\begin{center}
{\bf Acknowledgments}
\end{center}

S. S. thanks to Universidad de Chile, Facultad de Ciencias F\'{\i}sicas y Matem\'{a}ticas,
Departamento de F\'{\i}sica for their hospitality where part of this work was done. Also
to Profess.. S. Codriansky and P. Cordero for stimulating discussions.

\pagebreak

\end{document}